\documentclass[12pt]{iopart}
\usepackage{graphics}
\def\prl{{\em Phys. Rev. Lett. }}
\def\prc{{\em Phys. Rev. {\bf C} }}
\def\prd{{\em Phys. Rev. {\bf D} }}

\def\npa{{\em Nucl. Phys. {\bf A}}}

\def\epjc{{\em Eur. Phys. J. {\bf C}}}

\def\plb{{\em Phys. Lett. {\bf B}}}

\def\zpc{{\em Z. Phys. {\bf C}}}

\begin{document}
\title{Direct photons from relativistic heavy-ion collisions}
\author{Dinesh K. Srivastava\footnote{Email: dinesh@veccal.ernet.in}}
\address{Variable Energy Cyclotron Centre, 1/AF Bidhan Nagar, Kolkata 700064, 
India}

\begin{abstract}
We recall the seminal developments in the study of  radiation of direct photons
from relativistic heavy ion collisions, which have helped to
enhance the  scope of single photons as a probe of the quark gluon 
plasma considerably.  There is a
mounting evidence that in addition to providing information about the
initial temperature of the plasma as envisaged originally, these radiations  
measure the momentum anisotropy of the deconfined quarks and gluons,
energy loss of the quarks, the initial spatial asymmetry 
of the plasma, and  the history of evolution of the system.
After a brief description of the theoretical developments
and results for direct photons at SPS 
energies, we discuss the expectations and findings at RHIC energies.

\end{abstract}


%
\section{Introduction}

Relativistic collisions of heavy nuclei are being studied with the aim
of producing quark-gluon plasma, a deconfined state of strongly interacting
matter, which populated the early universe a few micro-seconds after the
big bang. Electromagnetic radiations in general, and direct 
photons in particular, from such collisions
are expected to provide an accurate information about the initial conditions
and the history of evolution of the plasma while it cools and hadronizes.
This is possible because, the photons interact only 
electromagnetically and therefore they have a  mean free path which
is much larger than the size of the system. Thus
 once produced, they leave the system without any re-interaction and
carry the information about the circumstances of their 
production~\cite{Feinberg:1976ua}.
This expectation has led to an intense and concerted effort
towards identification of various sources of such radiations- theoretically,
and to their
isolation from a background of intense radiations from the decay of
hadrons- experimentally. While initially these studies were aimed at
providing the initial temperature of the plasma, recent investigations
have thrown open several new and unique possibilities. It is now known that
single photons probe the evolution of the system size by
intensity interferometry\cite{int} as well as the evolution of the 
momentum anisotropy of quarks and gluons by the elliptic flow\cite{v2}.

  A new and vastly clean source of photons\cite{fms}, due to the
passage of high energy quarks through the quark gluon plasma has also
been suggested, which promises to provide an independent and accurate
 check on jet-quenching and several other aspects of the collision dynamics.
It is not possible to cover the details of these developmentns
in the limited space and time available to us, and the reader may
consult several excellent reviews~\cite{review} available by now.
\begin{figure}
\begin{center}
\resizebox{0.80\columnwidth}{!}{%
\includegraphics{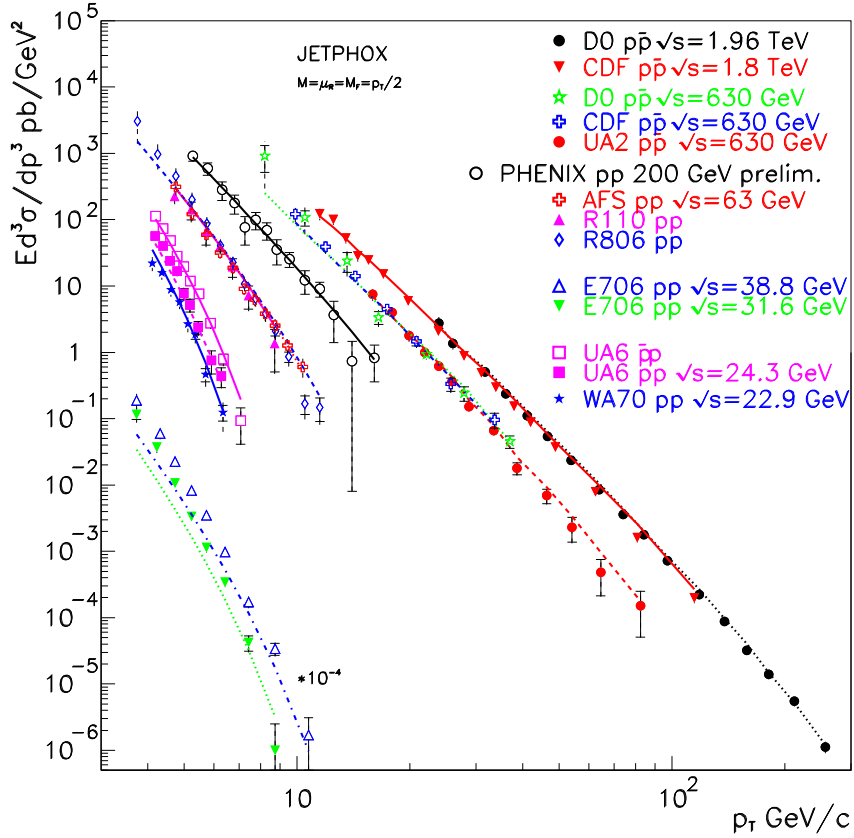}
}
\caption{World's inclusive and isolated direct photon production cross-section
measured in $ppi$ and $\overline{p}p$ collisions compared to
NLO predictions, using JETPHOX. BFG II (CTEQ6M) fragmentation (structure)
functions and a common scale $p_T/2$ were used. (Taken from Ref.\cite{pat})}.
\label{fig1}
\end{center}
\end{figure}

Direct photons or single photons are photons which do not have their
origin in the decay of hadrons, e.g., $\pi^0 \rightarrow \gamma \gamma$ 
or $\eta \rightarrow \gamma \gamma$, which account for up to 90--98\% of 
all the photons which are detected. These decays pose a serious experimental
challenge to the detection of direct photons and several methods, basically 
employing invariant mass analysis along with  statistical techniques of
mixed-events,  
have been utilized to subtract the decay photons from the inclusive
photon spectra. It should be recalled that suppression of high $p_T$ hadrons
due to jet-quenching seen at higher centre of mass energies leads to an
 increased  value for the ratio, $\gamma/\pi^0$. This considerably 
enhances the ease of
 detection of single photons at these momenta.  Let us briefly recall
the nodal theoretical developments in this field which have made these
studies so very rewarding.

\section{Recent Theoretical Developments}

Theoretically, we need to identify various sources of direct photons 
and understand their relative importance and  characteristics. 
One critically studied
source is the radiation of prompt photons due to Compton 
scattering ($q g \rightarrow q \gamma$) and annihilation 
($q \overline{q} \rightarrow g \gamma$) of partons of the colliding
 nucleons.  Perturbative QCD
calculations at the NLO are available~\cite{pat,vogel} and all the available
data for $pp$ and $p\overline{p}$ collisions have been analyzed, with the
inclusion of the fragmentation off the final state quarks
($q \rightarrow q \gamma$), and a quantitative description is obtained 
by choosing a scale such that the scales for factorization, renormalization,
and fragmentation are equal to $p_T/2$ (see Fig.\ref{fig1}). 
\begin{figure}
\begin{center}
\resizebox{0.80\columnwidth}{!}{%
\includegraphics{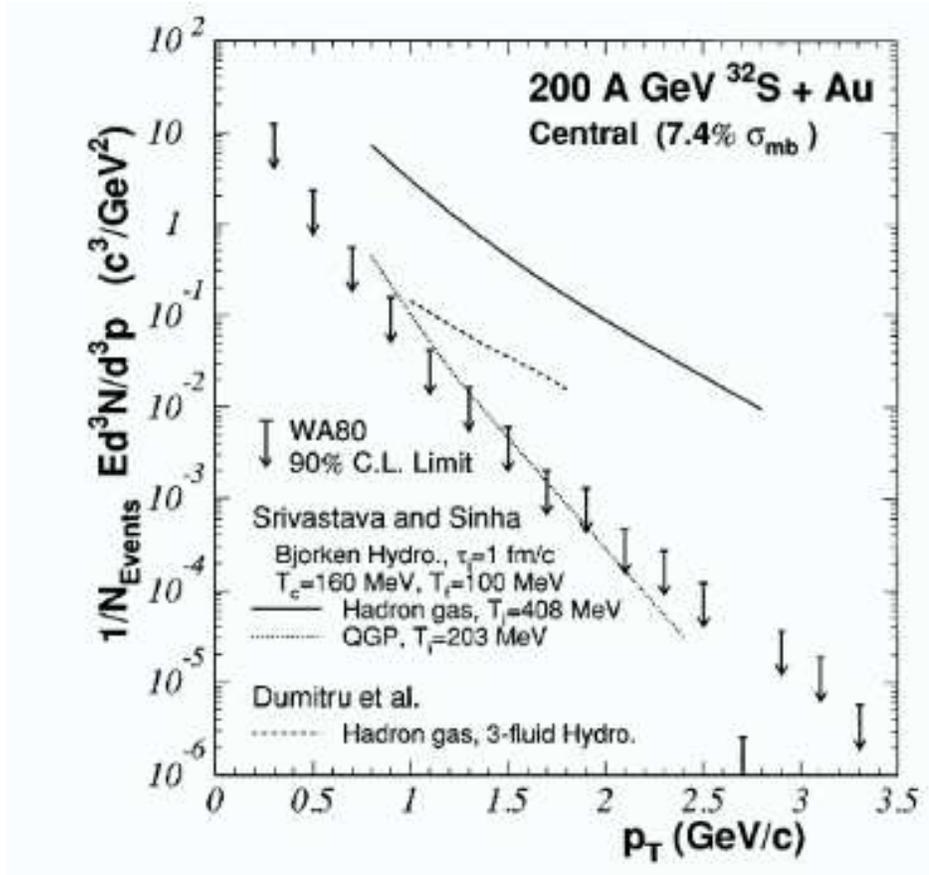}
}
\caption{
Upper limits at the 90\% confidence level
on the invariant excess photon yield for the 7.4\% $\sigma_{\mathrm mb}$
most central collisions of 200A GeV $^{32}S+Au$. The solid 
and the dashed curves give the thermal photon production 
expected from hot hadron gas calculations\cite{ss_80} while the 
dotted curve is the calculated thermal photon production
expected in the case of a QGP formation~\cite{ss_80}.(Taken from
Ref.\cite{wa80}.)
}
\label{fig2}
\end{center}
\end{figure}
Extension of these
results to the case of  nucleus-nucleus collisions requires care. Early studies 
often obtained the prompt photon contribution for them by scaling the
results for  $pp$ collisions for the corresponding $\sqrt{s}$ 
with the number of
nucleon-nucleon collisions. One immediately sees that this will over-estimate
the prompt photon production for nucleus-nucleus collisions  
as it ignores the iso-spin of the nucleons; protons ($uud$) and
neutrons ($udd$) have different number of up and down valence quarks.
Thus it will strongly affect the results for those
values of $p_T$ which derive a large contribution from the valence quarks.
On the other hand, there is no direct experimental method to get prompt photon 
contributions in $pn$ and $nn$ scatterings, though in principle one could
estimate them by comparing results with scatterings involving deuterons.
 Secondly, one has to account for the  effect of shadowing on structure
 functions. And lastly
and equally importantly, one has to account for the energy loss suffered
by final state quarks before they fragment into photons, if the collision
leads to formation of a quark-gluon plasma~\cite{jeon,arleo}. 
\begin{figure}
\begin{center}
\resizebox{0.50\columnwidth}{!}{%
\includegraphics{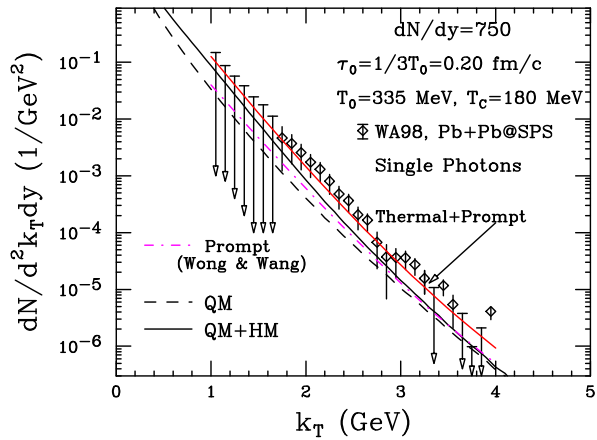}
}
\caption{
Single photon production in $Pb+Pb$ collision at the
CERN SPS. QM stands for radiations from the quark matter in the
QGP phase and the mixed phase, HM denotes the
radiation from the hadronic matter in the 
mixed phase and the hadronic phase. Prompt
photons are estimated using pQCD (with a K-factor
estimated using NLO calculations) and intrinsic
$k_T$ of partons. (Taken from Ref.\cite{ss_98}).
}
\label{fig3}
\end{center}
\end{figure}

Of-course the importance of single photons having intermediate $p_T$ lies
in their being dominated by thermal photons which have their origin in the
Compton and annihilation processes in the quark-gluon 
plasma\cite{joe,baier} as well as in the reactions in the
 hadronic phase following the hadronization~\cite{joe}.  
The calculations for radiation of photons from QGP have been extended up to
two loops~\cite{pat2}, which open up additional processes of
bremsstrahlung and annihilation of a off-shell quark with an anti-quark
following a scattering.  In a very significant and novel development,
complete leading order results are now
available~\cite{amy1}, which also properly account for the Landau
Pomeranchuk Migdal (LPM) suppression in the plasma. 

The production of photons due to hadronic reactions~\cite{joe} has also been 
studied extensively, using field theoretical models along with an
effective Lagrangian incorporating various mesonic fields and even baryons.
Results are available which additionally account for the likely medium
modification of hadronic properties~\cite{jane}. The most recent calculations
incorporate the strange mesons, using a Massive Yang-Mills theory, 
account for the form-factors, and include hitherto ignored $t$-channel
exchange of $\omega$ mesons which makes a large contribution at intermediate
$p_T$ and explore the consequences of including photon producing reactions
involving baryons~\cite{gale}. In view of these developments, it is quite
appropriate to suggest that the calculations of the rate of production of
photons have reached an unprecedented level of sophistication. 
\begin{figure}
\begin{center}
\resizebox{0.50\columnwidth}{!}{%
\includegraphics{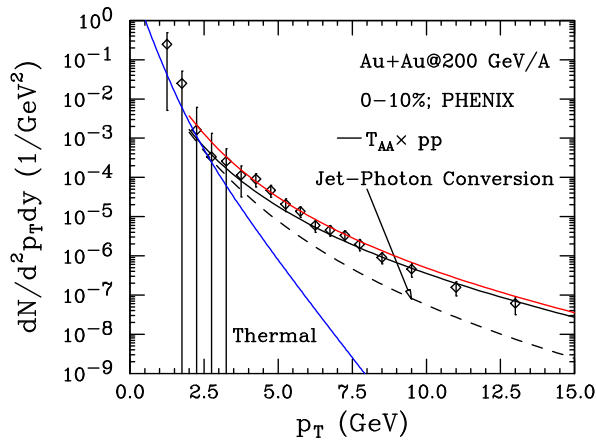}
}
\caption{
Yield of photons in central Au+Au collisions at $\sqrt{s}$ = 200 AGeV. The 
primary hard photons, jet-photon conversion and the sum of the
two can be seen here, indicating the presence of the later source. 
The thermal photon contribution is also given.
Data are from the PHENIX collaboration~\cite{auau_200}. (Taken from 
Ref.\cite{fms2})
}
\label{fig4}
\end{center}
\end{figure}

The other significant development involves an enormous progress made
 towards the understanding of production of photons due to passage of jets
through quark-gluon plasma, discovered recently~\cite{fms}. These can make 
a large contribution at $p_T \approx$ 4--8 GeV, at RHIC and LHC energies
where jet-quenching appears. It has also been realized that there would be
an additional contribution due to the jet-induced bremsstrahlung 
process~\cite{rus} in the plasma.
 A high level of sophistication was introduced by treating jet-quenching
and jet-induced photon production within the same treatment,
 which thus accounts
for the energy loss (and flavour change) suffered by the jets as they
traverse the plasma, and also treats the effect of this energy loss on 
the fragmentation contribution of the prompt photons as well 
as the jet-induced bremsstrahlung, with inclusion of LPM 
effects~\cite{amy2,gale2}.

\section{Results at SPS energies}

In order to get an idea of greatly increased insights  provided by single
photon production, let us briefly recall some of the important 
results from the SPS
era, many of which preceded the large strides made in
 our theoretical understanding mentioned above.
The first hint of single photon production, which later turned out to
be the upper limit of their production came from the S+Au collisions 
studied at the SPS energies~\cite{wa80}.

These results were analyzed in two different scenarios by authors of 
Ref.~\cite{ss_80}. In the first scenario, a thermally and chemically 
equilibrated quark-gluon plasma was assumed to be formed at some initial time
($\tau_0\approx$ 1 fm/$c$), which expanded~\cite{prd}, cooled, and converted
into a mixed phase
 of hadrons and QGP
at a phase-transition temperature, $T_C \approx$ 160 MeV. 
When all the quark-matter was 
converted into a hadronic matter, the hot hadronic gas continued to cool
and expand, and underwent a freeze-out at a temperature of about 140 MeV.
The hadronic gas was assumed to consist of $\pi$, $\rho$, $\omega$, and $\eta$
mesons, again in a thermal and chemical equilibrium.
 This was motivated by the fact that the included hadronic reactions
involved~\cite{joe} these mesons. This was already a considerable improvement
over a gas of mass-less pions used in the literature at that time.
 In the second scenario,
the collision was assumed to lead to a hot hadronic gas of the same composition.
The initial temperature was determined by  
demanding that the entropy of the system be determined from the measured
particle rapidity density~\cite{hwa}. It
was found that the scenario which did not involve a formation of QGP led to a
much larger initial temperature and a production of 
photons which was considerably larger than the upper limit of the photon
production, and could be ruled out. The calculation assuming a 
quark-hadron phase transition yielded results which were consistent with the
upper limit of the photon production. These results were 
confirmed~\cite{wa80_ana} by several calculations exploring different
models of expansion (see Fig.\ref{fig2}).
\begin{figure}
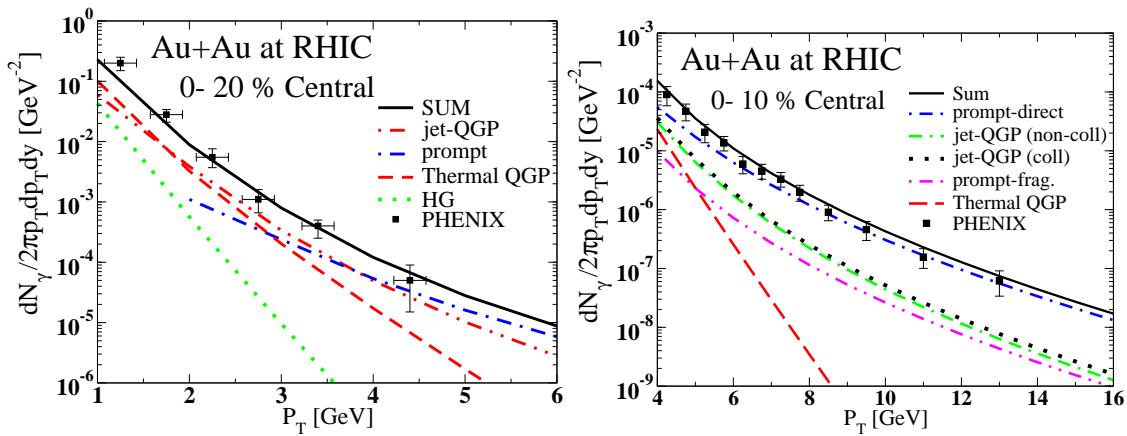

\begin{center}
\resizebox{0.95\columnwidth}{!}{%
\includegraphics{gale1.eps}
\includegraphics{gale2.eps}
}
\caption{
Yield of photons in Au+Au collisions at RHIC for two centrality
classes: 0-20\% (left panel) and  0-10\% (right panel).
Relative importance of various sources is seen clearly.
(Taken from Ref.\cite{gale2}, which may be 
referred for details).
Data are
from the PHENIX collaboration\cite{auau_200,auau_200_2}. 
}
\label{fig5}
\end{center}
\end{figure}

It was soon realized that one may not limit the hadronic gas to contain
just $\pi$, $\rho$, $\omega$, and $\eta$ mesons, as there was increasing
evidence that perhaps all the mesons and baryons were being produced in
a thermal and chemical equilibrium in such collisions. Thus
authors of Ref.~\cite{crs} explored the consequences of using a hadronic
gas consisting of essentially all the hadrons in the Particle Data Book,
in a thermal and chemical equilibrium. This led to an interesting result for
the Pb+Pb collision at SPS energies, for which experiments were in progress.
It was found that with the rich hadronic gas, the results for the production
of photons in the phase-transition and no phase transition models discussed
above were quite similar, suggesting that measurement of photons at the
SPS energy could perhaps not distinguish between the two cases. However,
in a very important observation, it was also noted that the calculations
involving hot hadronic gas at the initial time would lead to hadronic densities
of several hadron/fm$^3$, and while those involving a quark gluon plasma
in the initial state would be free from this malady. Thus, it was
concluded that the calculations
involving a phase transition to QGP offered a more natural description.

The WA98 experiment~\cite{wa98}, reported the first observation of direct 
photons in central 158A GeV Pb+Pb collisions studied at the CERN SPS.
This was explained~\cite{ss_98} in terms of formation of quark gluon plasma
in the initial state (at $\tau_0\approx$ 0.2 fm/$c$), which expanded,
 cooled and hadronized as in 
Ref.~\cite{crs} (see Fig.\ref{fig3}). An independent confirmation of this
approach was provided by an accurate description~\cite{yul} of excess 
dilepton spectrum measured by the NA60 experiment for the same system.

 Once again the results for single photons 
 were analyzed by several authors using varying models
of expansion as well as rates for production of photons; viz., with or with-out
medium modification of hadronic properties (see, 
e.g.,Ref.~\cite{jane2,pasi,gale}).  
The outcome of all these efforts can be summarized as follows: the single 
photon production in Pb+Pb collisions at SPS energies can be described
either by assuming a formation of QGP in the initial state or by assuming
the formation of a hot hadronic gas whose constituents have massively
modified properties. The later description, however, involved  a hadronic
density of several hadrons/fm$^3$, which raises doubts about the applicability
of a description in terms of hadrons, as suggested by Ref.~\cite{crs}.

\section{Results at RHIC energies: The light from quarks}

In view of the above, the results
for single photon production from Au+Au collisions at the highest RHIC
energy were eagerly awaited. The first to appear,
courtesy the suppression of hadrons due to jet-quenching, was the
centralitiy dependence of single photon production\cite{auau_200},
 where the single photons
were clearly identified for $p_T >$ 4 GeV 
for more central collisions and upper limits could be established
for $p_T$ down to about 2 GeV. The authors compared these
results with NLO pp values scaled by the number of collisions and reported
a good agreement. This result is sometimes used to suggest an absence of new
sources of single photons in this $p_T$ range. This approach, 
as discussed earlier, neglects the
iso-spin of the nucleons, which leads to a reduction of single photon
production at larger $p_T$, as well as the energy loss
suffered by final state quarks, befroe they fragment.
 Thus the above agreement, in fact, points to 
the presence of additional sources of single photons!

A first attempt to understand these data was made by authors of 
Ref.\cite{fms2}, who included prompt photons, thermal photons-
assuming a production of quark gluon plasma, {\em and} jet-conversion
photons due to passage of jets through the QGP.
The initial temperature (at $\tau_0 \approx$ 0.2 fm/$c$) was obtained 
by relating the entropy to the multiplicity density, measured
experimentally~\cite{hwa}. A good description of the
data was obtained, providing the first indication of photons due to the
mechanism of jet-photon conversion\cite{fms} (see Fig.\ref{fig4}).

These results were next analyzed with jet-induced photons, first with
a longitudinal expansion of the plasma~\cite{amy2} and next with 
transverse expansion of the plasma~\cite{gale2}. These results, along with
clear direct photon excess measured at lower $p_T$ (Ref.\cite{auau_200_2}) 
are shown in Fig.\ref{fig5}, which also shows various components of the
single photons at large $p_T$, viz., thermal, prompt-direct (i.e., Compton+
annihilation), 
prompt-fragmentation (corrected for the energy loss suffered by the
quarks before they fragment), and the jet-conversion photons- both due
to the Compton and annihilation processes and the jet-induced bremsstrahlung
process, the later corrected for energy loss suffered by the jet. The
iso-spin and shadowing effects were also properly accounted for. These
results represent a very high level of sophistication reached in these
studies. The studies also indicate a substantial production of photons due
to these processes at LHC energies\cite{fms,gale2}
 and have been extended to production
of dileptons as well~\cite{gale3}.

In a related and independent attempt, single photon production has been 
calculated using the parton cascade model. It has been found that
large $p_T$ photon production at RHIC and even at SPS
 is quantitatively reproduced by these 
calculations provided the LPM effect on radiation of gluons is 
accounted for~\cite{pcm}. It has also been pointed out that the
study of photon (or dilepton~\cite{jet-dil}) tagged jets can help us
differentiate between different models for jet-quenching~\cite{renk}. 
\begin{figure}
\begin{center}
\resizebox{0.50\columnwidth}{!}{%
\includegraphics{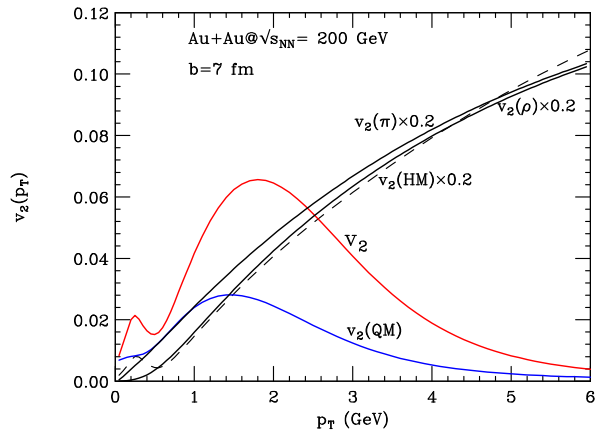}
}
\caption{
$v_2(p_T)$ for thermal photons from off-central Au+Au collisions at b=7 fm
for $\sqrt{s}$= 200 AGeV. Quark and hadronic matter contributions are shown 
separately. The elliptic flow of $\pi$ and $\rho$ mesons are shown for a
comparison. The peak at lower $p_T$ arises as the photon production due
to
the reaction $\pi \pi \rightarrow \rho \gamma$ loses out to that due to  
 $\pi \rho \rightarrow \pi \gamma$, as $p_T$ increases.
(Taken from Ref.\cite{v2}, which may be 
referred for details).
}
\label{fig6}
\end{center}
\end{figure}

A very important and useful information about the evolution of
the momentum anisotropy of deconfined quarks and gluons in such collisions
 can be obtained by the
measurement of the effect of the elliptic
flow on thermal photons or dileptons\cite{v2}. In a very interesting
confirmation of our models for the evolution of the plasma~\cite{kolb},
 the $v_2$ for
thermal photons rises with $p_T$, reaches a maximum, and then falls, when
the contributions from the QGP phase dominate, where the temperatures 
are large but the flow is still building up (see Fig.\ref{fig6}).
 An experimental verification~\cite{miki}
of these results would provide a direct confirmation of the production
of QGP at very early times in such collisions. 
A small azimuthal anisotropy  (with negative $v_2$) in the momentum
 distribution of photons at 
large $p_T$ is also expected, which reflects the spatial anisotropy
of the plasma, at the earliest times in non-central collisions~\cite{gale2}.

Before concluding, one may recall a very valuable comparison of
several calculations describing the low $p_T$ single photon production from
Au+Au collisions at 200A GeV, which all point, rather strongly,
to the presence of a strong
radiation from quark-gluon plasma produced very early in the 
collision\cite{david}, which has a temperature of about 300--500 MeV at
$\tau_0$ of about 0.6--0.2 fm/$c$. 

\section{Summary} We conclude that the study of single photons has opened
up very many interesting possibilities to explore the initial state of the
quark-gluon plasma and its evolution.  The light from quarks
is clearly seen at RHIC energies! 

These results hold out a 
promise~\cite{review} of a 
spectacular display of all the aspects discussed here at the LHC energies.

\ack
The author gratefully acknowledges a very enjoyable and rewarding
collaboration spanning almost two decades
with Terry Awes, Steffen Bass, Rupa Chatterjee, Jean Cleymans, Rainer Fries,
 Evan Frodermann, Charles Gale, Ulrich Heinz, Berndt M\"uller, 
Thorsten Renk, Krzysztof Redlich, Bikash Sinha,
and Simon Turbide, and valuable discussions with countless colleagues.

\end{document}